    \documentclass[aps,pre,groupedaddress,amsmath,eqsecnum,showpacs]{revtex4}
    \usepackage{bm}
    \usepackage{graphicx}
\newcommand\beq{\begin{equation}}
\newcommand\eeq{\end{equation}}
\newcommand\beqa{\begin{eqnarray}}
\newcommand\eeqa{\end{eqnarray}}
\newcommand{\dd}{\text{d}}
\newcommand{\nn}{\nonumber\\}

\newcommand{\qq}{r}

\newcommand{\al}{\alpha}
\newcommand{\at}{\frac{1+\alpha}{2}}
\newcommand{\atp}{\left(\frac{1+\alpha}{2}\right)}
\newcommand{\sg}{(\widehat{\bm{\sigma}}\cdot\mathbf{g})}
\newcommand{\si}{\widehat{\sigma}}
\newcommand{\nutz}{\nu_{2|0}}
\newcommand{\nuzt}{\nu_{0|2}}
\newcommand{\nuto}{\nu_{2|1}}
\newcommand{\nuzth}{\nu_{0|3}}
\newcommand{\nufz}{\nu_{4|0}}
\newcommand{\nutt}{\nu_{2|2}}
\newcommand{\nuzf}{\nu_{0|4}}

\newcommand{\omegazt}{\omega_{0|2}}
\newcommand{\omegato}{\omega_{2|1}}
\newcommand{\omegazth}{\omega_{0|3}}
\newcommand{\omegafz}{\omega_{4|0}}
\newcommand{\omegatt}{\omega_{2|2}}
\newcommand{\omegazf}{\omega_{0|4}}

\newcommand{\omeganu}{\nu}

\begin{document}
\title{Third and fourth degree collisional moments for inelastic Maxwell models}
\author{Vicente Garz\'o}
\email{vicenteg@unex.es}
\homepage{http://www.unex.es/fisteor/vicente/}
\author{Andr\'es Santos}
\email{andres@unex.es} \homepage{http://www.unex.es/fisteor/andres/}
\affiliation{Departamento de F\'{\i}sica, Universidad de
Extremadura, E-06071 Badajoz, Spain}
\date{\today}
\begin{abstract}
The third and fourth degree collisional moments for $d$-dimensional
inelastic Maxwell models are exactly evaluated in terms of the
velocity moments, with explicit expressions for the associated
eigenvalues and cross coefficients as functions of the coefficient
of normal restitution. The results are applied to the analysis of
the time evolution of the moments (scaled with the thermal speed) in
the free cooling problem. It is observed that the characteristic
relaxation time toward the homogeneous cooling state decreases as
the anisotropy of the corresponding moment increases. In particular,
in contrast to what happens in the one-dimensional case, all the
anisotropic moments of degree equal to or less than four vanish in
the homogeneous cooling state for $d\geq 2$.
\end{abstract}

\maketitle

\maketitle

\section{introduction}
A realistic model capturing the influence of dissipation on the dynamic properties of granular systems consists
of a gas of inelastic hard spheres (IHS) with a constant coefficient of normal restitution $\alpha \leq 1$
\cite{BP04}. For  sufficiently low densities, the Boltzmann equation for IHS provides the adequate framework to
describe the time evolution of the one-particle velocity distribution function $f(\mathbf{r},\mathbf{v};t)$
\cite{BDS97}. However, the intricacy of the Boltzmann collision operator for IHS makes it difficult to obtain
exact results. For instance, the fourth cumulant $a_2$ of the velocity distribution in the so-called homogeneous
cooling state (HCS) is not exactly known, although good estimates of it have been proposed
\cite{NE98,MS00,BP06}. For inhomogeneous situations, explicit expressions for the Navier--Stokes (NS) transport
coefficients are approximately obtained by considering the leading terms in a Sonine polynomial expansion
\cite{BDKS98,BC01,GD02,GSM07}.

As in the elastic case, part of the above difficulties can be overcome by considering the so-called Maxwell
models, i.e., models for which the collision rate is independent of the relative velocity of the two colliding
particles. Inelastic Maxwell models (IMM) have attracted the attention of physicists and mathematicians since
the beginning of the century
\cite{BCG00,CCG00,NK00,C01,EB02a,EB02,EB02bis,BMP02,KN02,KN02bis,NK02,BC02,MP02,MP02bis,BC03,BCT03,SE03,S03,
NK03,G03,BE04,GA05,BG06,ETB06,S07}. The structure of the Boltzmann collision operator for IMM has the advantage
of allowing for the derivation of a number of exact properties, such as the high-velocity tails
\cite{EB02a,EB02,EB02bis,BMP02,KN02bis,KN02} and the cumulants \cite{BMP02,EB02,MP02,MP02bis,S03,GA05} in
homogeneous situations, the NS transport coefficients \cite{S03,GA05}, and the rheology under simple shear flow
\cite{C01,G03}. As a consequence, it is possible to explore the influence of inelasticity on the dynamic
properties in a clean way, without the need of introducing additional, and sometimes uncontrolled,
approximations. Apart from their academic interest, it turns out that the IMM  reliably describe the properties
of IHS in some situations, as happens in the simple shear flow problem \cite{G03} and for the NS transport
coefficients associated with the mass flux \cite{GA05}. Furthermore, it is interesting to remark that recent
experiments \cite{KSSAON05} for magnetic grains with dipolar interactions are well described by IMM.

The aim of this paper is to contribute to the advancement in the knowledge of exact properties of IMM by
evaluating all the third and fourth degree moments of the Boltzmann  collision operator for an arbitrary number
of dimensions $d$. The knowledge of those collisional moments, along with that of the second degree collisional
moments \cite{S03,G03}, opens up a number of interesting applications. For instance, one can investigate the
temporal relaxation toward the HCS, starting from arbitrary initial conditions (not necessarily isotropic), as
measured by the lowest degree moments (namely, the fourth degree moments) which signal the non-Gaussian
character of the asymptotic velocity distribution function. This issue will be covered in this paper.

This paper is organized as follows. In Section \ref{sec2} the Boltzmann equation for IMM is presented. Next, the
Ikenberry polynomials \cite{TM80} $Y_{2r|i_1i_2\ldots i_s}(\mathbf{V})$ of degree $k=2r+s$ are introduced and
their associated collisional moments $J_{2r|i_1i_2\ldots i_s}$ for $k=3$ and 4 are evaluated, the technicalities
being relegated to an Appendix. The results are applied to the relaxation problem of the (scaled) moments toward
their asymptotic values in the HCS in Section \ref{sec3}. The paper is closed in Section \ref{sec4} with a brief
discussion of the results obtained here.

\section{Collisional moments for IMM\label{sec2}}

In the absence of external forces, the inelastic Boltzmann equation for a granular gas reads \cite{BP04}
 \beq
\left(\partial_t +\mathbf{v}\cdot\nabla\right)f(\mathbf{r},\mathbf{v};t)=J[\mathbf{v}|f,f], \label{2.1}
\eeq
where $J[\mathbf{v}|f,f]$ is the Boltzmann collision operator. The form of the operator $J$ for IMM can be
obtained from the form for IHS by replacing the IHS collision rate (which is proportional to the relative
velocity of the two colliding particles) by an effective velocity-independent collision rate. With this
simplification, the form of $J$ becomes \cite{NK03}
\begin{equation}
J\left[{\bf v}_{1}|f,f\right] =\frac{\omeganu}{n\Omega_d} \int
\dd{\bf v}_{2}\int \dd\widehat{\boldsymbol{\sigma}} \left[
\alpha^{-1}f({\bf v}_{1}')f({\bf v}_{2}')-f({\bf v}_{1})f({\bf
v}_{2})\right] .
\label{1}
\end{equation}
Here,
\beq
n=\int\dd \mathbf{v} f(\mathbf{v})
\eeq
is the number density, $\omeganu$ is an effective collision
frequency (to be chosen later), $\Omega_d=2\pi^{d/2}/\Gamma(d/2)$ is
the total solid angle in $d$ dimensions, and $\alpha\leq 1$ refers
to the constant coefficient of restitution. In addition, the primes
on the velocities denote the initial values $\{{\bf v}_{1}^{\prime},
{\bf v}_{2}^{\prime}\}$ that lead to $\{{\bf v}_{1},{\bf v}_{2}\}$
following a binary collision:
\begin{equation}
\label{3}
{\bf v}_{1}^{\prime}={\bf v}_{1}-\frac{1}{2}\left( 1+\alpha
^{-1}\right)(\widehat{\boldsymbol{\sigma}}\cdot {\bf
g})\widehat{\boldsymbol {\sigma}}, \quad {\bf v}_{2}^{\prime}={\bf
v}_{2}+\frac{1}{2}\left( 1+\alpha^{-1}\right)
(\widehat{\boldsymbol{\sigma}}\cdot {\bf
g})\widehat{\boldsymbol{\sigma}}\;,
\end{equation}
where ${\bf g}={\bf v}_1-{\bf v}_2$ is the relative velocity of the colliding pair and
$\widehat{\boldsymbol{\sigma}}$ is a unit vector directed along the centers of the two colliding spheres. The
collision frequency $\omeganu$ can be seen as a free parameter in the model. Its dependence on the coefficient
of restitution $\alpha$ can be chosen to optimize the agreement with  the results obtained from the Boltzmann
equation  for IHS. In particular, to get the same expression for the cooling rate as the one found for IHS
(evaluated in the local equilibrium approximation) one takes the choice \cite{S03}
\begin{equation}
\label{4}
\omeganu=\frac{d+2}{2}\nu_0, \quad
\nu_0=\frac{4\Omega_d}{\sqrt{\pi}(d+2)}n\sigma^{d-1}\sqrt{\frac{T}{m}},
\end{equation}
where $\sigma$ is the diameter of the spheres. Note that, in any
case, the results derived in this paper will be independent of the
specific choice of $\nu_0$.

A useful identity for an arbitrary function $h({\bf v})$ is given by
\begin{equation}
\label{5}
\mathcal{J}[h]\equiv \int \dd{\bf v}_1 h({\bf v}_1)J[{\bf
v}_1|f,f]=\frac{\omeganu}{n\Omega_d} \int \dd{\bf v}_{1}\,\int
\dd{\bf v}_{2}f({\bf v}_{1})f({\bf v}_{2}) \int
\dd\widehat{\boldsymbol{\sigma}}\,\left[h({\bf v}_1'')-h({\bf
v}_1)\right],
\end{equation}
where
\begin{equation}
\label{6}
{\bf v}_{1}''={\bf v}_{1}-\frac{1}{2}(1+\alpha)(
\widehat{\boldsymbol{\sigma }}\cdot {\bf
g})\widehat{\boldsymbol{\sigma}}
\end{equation}
denotes the post-collisional velocity. If $h(\mathbf{v})$ is a
polynomial, then
\beq
\mathcal{M}[h]\equiv \int \dd\mathbf{v} h(\mathbf{v}) f(\mathbf{v})
\label{2.2}
\eeq
is its associated velocity moment and $\mathcal{J}[h]$ is the
corresponding collisional moment.

In the case of Maxwell models (both elastic and inelastic), it is convenient to introduce the Ikenberry
polynomials \cite{TM80} $Y_{2r|i_1i_2\ldots i_s}(\mathbf{V})$ of degree $k=2r+s$, where
$\mathbf{V}=\mathbf{v}-\mathbf{u}(\mathbf{r})$ is the peculiar velocity, $\mathbf{u}(\mathbf{r})$ being the mean
flow velocity defined as
\beq \mathbf{u}=\frac{1}{n}\int\dd \mathbf{v} \mathbf{v}f(\mathbf{v}). \eeq
The
Ikenberry polynomials are defined as $Y_{2r|i_1i_2\ldots i_s}(\mathbf{V})=V^{2r}Y_{i_1i_2\ldots
i_s}(\mathbf{V})$, where $Y_{i_1i_2\ldots i_s}(\mathbf{V})$ is obtained by subtracting from
$V_{i_1}V_{i_2}\ldots V_{i_s}$ that homogeneous symmetric polynomial of degree $s$ in the components of
$\mathbf{V}$ such as to annul the result of contracting the components of $Y_{i_1i_2\ldots i_s}(\mathbf{V})$ on
any pair of indices. The polynomials functions $Y_{2r|i_1i_2\ldots i_s}(\mathbf{V})$ of degree smaller than or
equal to four are \beq Y_{0|0}(\mathbf{V})=1,\quad Y_{0|i}(\mathbf{V})=V_i, \label{X0} \eeq \beq
Y_{2|0}(\mathbf{V})=V^2,\quad Y_{0|ij}(\mathbf{V})=V_iV_j-\frac{1}{d}V^2\delta_{ij}, \label{X1} \eeq \beq
Y_{2|i}(\mathbf{V})=V^2 V_i,\quad
Y_{0|ijk}(\mathbf{V})=V_iV_jV_k-\frac{1}{d+2}V^2\left(V_i\delta_{jk}+V_j\delta_{ik}+V_k\delta_{ij}\right),
\label{X2} \eeq \beq Y_{4|0}(\mathbf{V})=V^4,\quad
Y_{2|ij}(\mathbf{V})=V^2\left(V_iV_j-\frac{1}{d}V^2\delta_{ij}\right), \label{X3} \eeq \beqa
Y_{0|ijk\ell}(\mathbf{V})&=&V_iV_jV_kV_\ell-\frac{1}{d+4}V^2
\left(V_iV_j\delta_{k\ell}+V_iV_k\delta_{j\ell}+V_iV_\ell\delta_{jk}
+V_jV_k\delta_{i\ell}+V_jV_\ell\delta_{ik}+V_kV_\ell\delta_{ij}\right)\nn
&&+\frac{1}{(d+2)(d+4)}V^4\left(\delta_{ij}\delta_{k\ell}+\delta_{ik}\delta_{j\ell}+\delta_{i\ell}\delta_{jk}\right)\nn
&=&V_iV_jV_kV_\ell-\frac{1}{d+4}\left[Y_{2|ij}(\mathbf{V})\delta_{k\ell}+Y_{2|ik}(\mathbf{V})\delta_{j\ell}
+Y_{2|i\ell}(\mathbf{V})\delta_{jk} +Y_{2|jk}(\mathbf{V})\delta_{i\ell}\right.\nn &&\left.
+Y_{2|j\ell}(\mathbf{V})\delta_{ik}
+Y_{2|k\ell}(\mathbf{V})\delta_{ij}\right]-\frac{1}{d(d+2)}V^4\left(\delta_{ij}\delta_{k\ell}+
\delta_{ik}\delta_{j\ell}+\delta_{i\ell}\delta_{jk}\right). \label{X4} \eeqa

Here we will use the notation $M_{2r|i_1i_2\ldots
i_s}=\mathcal{M}[Y_{2r|i_1i_2\ldots i_s}]$ and $J_{2r|i_1i_2\ldots
i_s}=\mathcal{J}[Y_{2r|i_1i_2\ldots i_s}]$ for the associated
moments and collisional moments, respectively. Note that
$M_{0|0}=n$, $J_{0|0}=0$ (conservation of mass), $M_{0|i}=0$ (by
definition of the peculiar velocity), $J_{0|i}=0$ (conservation of
momentum), and $M_{2|0}=pd/m$, where $p=nT$ is the hydrostatic
pressure, $T$ being the granular temperature. Moreover,
$M_{0|ij}=(P_{ij}-p\delta_{ij})/m$, where $P_{ij}$ is the pressure
tensor and $M_{2|i}=2q_i/m$, where $\mathbf{q}$ is the heat flux
vector. The moment $M_{2|0}$,  the number density $n$, and the flow
velocity $\mathbf{u}$ are the hydrodynamic fields, while the moments
$M_{0|ij}$ and $M_{2|i}$ constitute the momentum and energy fluxes,
respectively. The remaining third degree moments $M_{0|ijk}$ and the
moments of a degree $k\geq 4$ are not directly related to the
hydrodynamic description, but they are useful to provide information
about the velocity distribution function. In particular, the moment
$M_{4|0}$ is related to the fourth cumulant $a_2$ as
\beq
a_2=\frac{m^2 }{d(d+2)nT^2}M_{4|0}-1,
\label{2.3}
\eeq
while the moments $M_{0|ijk}$, $M_{0|ijk\ell}$, and $M_{2|ij}$
measure the degree of anisotropy of the velocity distribution.

As in the elastic case, the mathematical structure of the collision
operator \eqref{1} implies that a collisional moment of degree $k$
can be expressed in terms of velocity moments of a degree less than
or equal to $k$. More specifically, the choice of the polynomials
$Y_{2r|\bar{s}}(\mathbf{V})$, where we have introduced the
short-hand notation $\bar{s}\equiv i_1i_2\ldots i_s$, yields the
following structural form for the collisional moments
$J_{2r|\bar{s}}$:
\beq
J_{2r|\bar{s}}=-\nu_{2r|s}
M_{2r|\bar{s}}+{\sum_{r',r'',\bar{s}',\bar{s}''}}^\dagger\lambda_{r'r''|\bar{s}'\bar{s}''\bar{s}}M_{2r'|\bar{s}'}M_{2r''|\bar{s}''},
\label{2.4}
\eeq
where the dagger in the summation denotes the constraints
$2(r'+r'')+s'+s''=2r+s$, $2r'+s'\geq 2$, and $2r''+s''\geq 2$. Since
the first term on the right-hand side of  Eq.\ \eqref{2.4} is
linear, then $\nu_{2r|s}$ represents the eigenvalue of the
linearized collision operator corresponding to the eigenfunction
$Y_{2r|\bar{s}}(\mathbf{V})$.

Let us now display the explicit expressions for the collisional
moments $J_{2r|i_1i_2\ldots i_s}$ for $k=2r+s\leq 4$. We start with
the second degree moments.

\subsection{Second degree collisional moments}
The second degree collisional moments were already evaluated in
Ref.\ \onlinecite{S03}. They are given by
\beq
J_{2|0}=-\nutz M_{2|0},\quad J_{0|ij}=-\nuzt M_{0|ij},
\label{X5}
\eeq
where the expressions for the eigenvalues $\nutz$ and $\nuzt$ are
\beq
\nutz=\frac{d+2}{4d}\left(1-\al^2\right)\nu_0,
\label{X6a}
\eeq
\beq
\nuzt=\frac{(1+\al)(d+1-\al)}{2d}\nu_0=\nutz+\frac{(1+\alpha)^2}{4}\nu_0.
\label{X6}
\eeq
 The quantity $\nutz$ is not but
the cooling rate, i.e., the rate of change of the granular
temperature due to the inelasticity of the collisions. The
eigenvalue   $\nuzt$ is the collision frequency associated with the
NS shear viscosity and  reduces to $\nu_0$ in the elastic limit. The
second equality in Eq.\ \eqref{X6} decomposes $\nuzt$ into the part
inherent to the collisional cooling plus the genuine part of the
momentum collisional transfer. As shown below, a similar
decomposition can be carried out for the eigenvalues $\nu_{2r|s}$.

\subsection{Third degree collisional moments}
The evaluation of the  third degree collisional moments $J_{2|i}$
and $J_{0|ijk}$ is performed in the Appendix. The results are
\beq J_{2|i}=-\nuto M_{2|i}, \quad J_{0|ijk}=-\nuzth M_{0|ijk},
\label{X7}
\eeq
where
\beq
\nuto=\frac{(1+\al)\left[5d+4-\al(d+8)\right]}{8d}\nu_0=\frac{3}{2}\nutz+\frac{(1+\al)^2(d-1)}{4d}\nu_0,
\label{X7b}
\eeq
\beq
 \nuzth=\frac{3}{2}\nuzt.
\label{X8}
\eeq
Equation \eqref{X7b} was first obtained in Ref.\ \onlinecite{S03}.
The eigenvalue $\nuto$ is the collision frequency associated with
the NS thermal conductivity. It is interesting to note that
$(\nuto-\frac{3}{2}\nutz)/(\nuzt-\nutz)=(d-1)/d$, which generalizes
the simple relationship, holding for elastic Maxwell models, between
the collision frequencies associated with the thermal conductivity
and the shear viscosity. An even simpler extension is provided by
Eq.\ \eqref{X8}.

\subsection{Fourth degree collisional moments}
The fourth degree collisional moments are also worked out  in the
Appendix. They can be written as
\beq
J_{4|0}=-\nufz M_{4|0}+\lambda_1 n^{-1}M_{2|0}^2-\lambda_2
n^{-1}M_{0|ij}M_{0|ji},
\label{X9}
\eeq
\beq
J_{2|ij}=-\nutt M_{2|ij}+\lambda_3 n^{-1}M_{2|0}M_{0|ij}-\lambda_4
n^{-1}\left(M_{0|ik}M_{0|kj}-\frac{1}{d}M_{0|k\ell}M_{0|\ell
k}\delta_{ij}\right),
\label{X11}
\eeq
\beqa
J_{0|ijk\ell}&=&-\nu_{0|4}M_{0|ijk\ell}+\lambda_5
n^{-1}\left[M_{0|ij}M_{0|k\ell}+
M_{0|ik}M_{0|j\ell}+M_{0|i\ell}M_{0|jk}-\frac{2}{d+4}\left(
M_{0|ip}M_{0|pj}\delta_{k\ell}\right.\right. \nn
&&\left.+M_{0|ip}M_{0|pk} \delta_{j\ell}
+M_{0|ip}M_{0|p\ell}\delta_{jk}+M_{0|jp}M_{0|pk}\delta_{i\ell}
+M_{0|jp}M_{0|p\ell}\delta_{ik}+M_{0|kp}M_{0|p\ell}\delta_{ij}\right)\nn
&&\left.
+\frac{2}{(d+2)(d+4)}M_{0|pq}M_{0|qp}\left(\delta_{ij}\delta_{k\ell}+
\delta_{ik}\delta_{j\ell}+\delta_{i\ell}\delta_{jk}\right) \right].
\label{X12}
\eeqa
In Eqs.\ (\ref{X9})--(\ref{X12}),  the usual summation convention
over repeated indices is assumed. The collision frequencies (or
eigenvalues) $\nu_{2r|{s}}$ and the cross coefficients $\lambda_i$
are given by
\begin{figure}
\includegraphics[width=0.95\columnwidth]{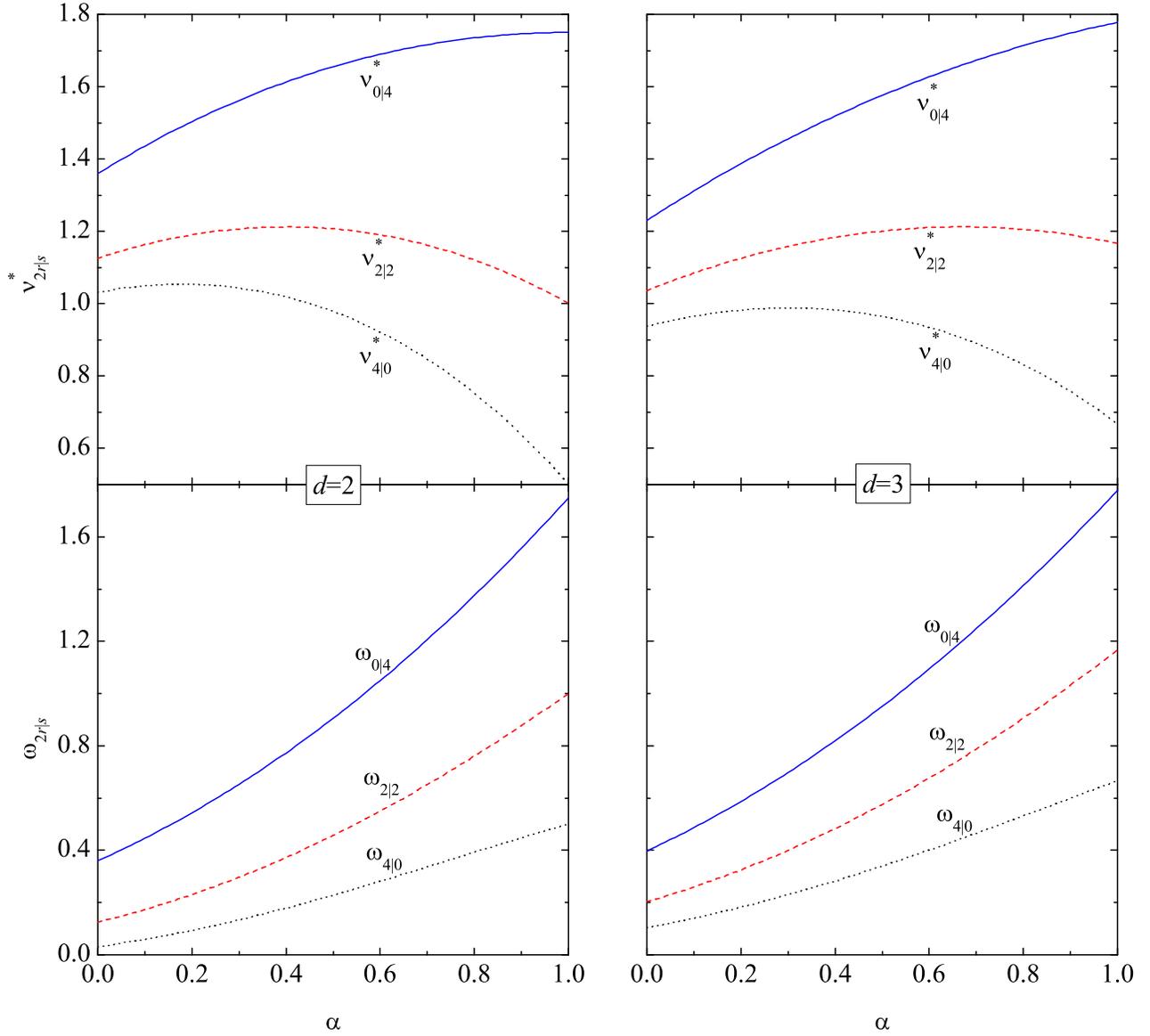}
\caption{(Color online) Plot of the (reduced) fourth degree
eigenvalues $\nuzf^*$, $\nutt^*$, and $\nufz^*$ (top panels) and of
the shifted eigenvalues $\omegazf$, $\omegatt$, and $\omegafz$
(bottom panels) as functions of the coefficient of restitution. The
left and right panels correspond to $d=2$ and $d=3$, respectively.
\label{fig1}}
\end{figure}
\beqa
\nufz&=&\frac{(1+\al)\left[12d+9-\alpha(4d+17)+3\alpha^2-3\alpha^3\right]}{16d}\nu_0\nn
&=&2\nutz+\frac{(1+\al)^2\left(4d-7+6\alpha-3\alpha^2\right)}{16d}\nu_0,
\label{X10}
\eeqa
\beqa
\nutt&=&\frac{(1+\al)\left[7d^2+31d+18-\al(d^2+14d+34)
+3\al^2(d+2)-6\al^3\right]}{8d(d+4)}\nu_0\nn &=&
2\nutz+\frac{(1+\al)^2\left[3d^2+7d-14+3\al(d+4)-6\al^2\right]}{8d(d+4)}\nu_0,
\label{X15}
\eeqa
\beqa
\nuzf&=&\frac{(1+\al)\left[2d^3+21d^2+61d+
39-3\al(d+3)(d+5)+3\al^2(d+3)-3\al^3\right]}{2d(d+4)(d+6)}\nu_0\nn
&=&2\nutz+\frac{(1+\al)^2\left[d^3+9d^2+17d-9+3\al(d+4)-3\al^2\right]}{2d(d+4)(d+6)}\nu_0,
\label{X18}
\eeqa
\beq
\lambda_1=\frac{(1+\al)^2(d+2)\left(4d-1-6\al+3\al^2\right)}{16d^2}\nu_0,
\label{X14}
\eeq
\beq
\lambda_2=\frac{(1+\al)^2\left(1+6\al-3\al^2\right)}{8d}\nu_0,
\label{X13}
\eeq
\beq
\lambda_3=\frac{(1+\al)^2\left[d^2+5d-2-3\al(d+4)+6\al^2\right]}{8d^2}\nu_0,
\label{X17}
\eeq
\beq
\lambda_4=\frac{(1+\al)^2\left[2-d+3\al(d+4)-6\al^2\right]}{4d(d+4)}\nu_0,
\label{X16}
\eeq
\beq
\lambda_5=\frac{(1+\al)^2\left[d^2+7d+9-3\al(d+4)+3\al^2\right]}{2d(d+4)(d+6)}\nu_0.
\label{X19}
\eeq
Equations \eqref{X10} and \eqref{X14} coincide with the results of
Ref.\ \onlinecite{S03}.

We have checked that, in the elastic case ($\alpha=1$) and for three-dimensional systems ($d=3$), all the
expressions reported in this Section reduce to known results \cite{TM80,GS03}. In the one-dimensional elastic
case ($d=1$, $\alpha=1$), the gas behaves as an ideal gas because a collision is equivalent to exchanging the
labels of both colliding particles. As a consequence, $J_{2r|\bar{s}}=0$. It is easy to check that Eqs.\
\eqref{X5}--\eqref{X19} are consistent with this property since the coefficients affecting the non-vanishing
moments are zero, i.e., $\nuto=\nufz=\lambda_1=0$.
 Moreover, for  one-dimensional inelastic gases ($d=1$, $\alpha<1$),
our expressions for $\nutz$, $\nuto$, $\nufz$, and $\lambda_1$ agree with the results derived by Ben-Naim and
Krapivsky \cite{NK00}, who obtained the exact expressions for all the collisional moments, namely $J_{2r|0}$ and
$J_{2r|x}$.

While the $\alpha$ dependence of the second and third degree
eigenvalues, Eqs.\ \eqref{X6a}, \eqref{X6}, \eqref{X7b}, and
\eqref{X8}, is relatively simple, that of the fourth degree
eigenvalues  \eqref{X10}--\eqref{X18} and the cross coefficients
\eqref{X14}--\eqref{X19} is more involved. Figure \ref{fig1} shows
the $\alpha$ dependence of the (reduced) eigenvalues $\nufz^*$,
$\nutt^*$, and $\nuzf^*$, where
$\nu_{2r|s}^*\equiv\nu_{2r|s}/\nu_0$, and the shifted eigenvalues
$\omegafz$, $\omegatt$, and $\omegazf$, where we have called
$\omega_{2r|s}\equiv \nu_{2r|s}^*-(r+s/2)\nutz^*$, for $d=2$ and
$d=3$. While $\nuzf^*$ decays monotonically as the inelasticity
increases, the other two eigenvalues $\nutt^*$ and $\nufz^*$ start
growing, reach a maximum, and then decay. The maximum  value of
$\nutt^*$  occurs at $\alpha\simeq 0.40$ for $d=2$ and at
$\alpha\simeq 0.67$ for $d=3$. In the case of $\nufz^*$, the maximum
occurs at $\alpha\simeq 0.18$ and $\alpha\simeq 0.30$ for $d=2$ and
$d=3$, respectively. However, when the part associated with the
cooling rate is subtracted from the bare eigenvalues, the resulting
shifted quantities $\omegafz$, $\omegatt$, and $\omegazf$ exhibit a
monotonic behavior. As shown in Section \ref{sec3}, these shifted
eigenvalues are the relevant ones in the time relaxation  of the
\emph{scaled} moments in the free cooling problem. Therefore, the
decrease of $\omegafz$, $\omegatt$, and $\omegazf$ implies that the
characteristic relaxation times of the (scaled) fourth degree
moments toward their asymptotic values increase with dissipation.

\begin{figure}
\includegraphics[width=0.95\columnwidth]{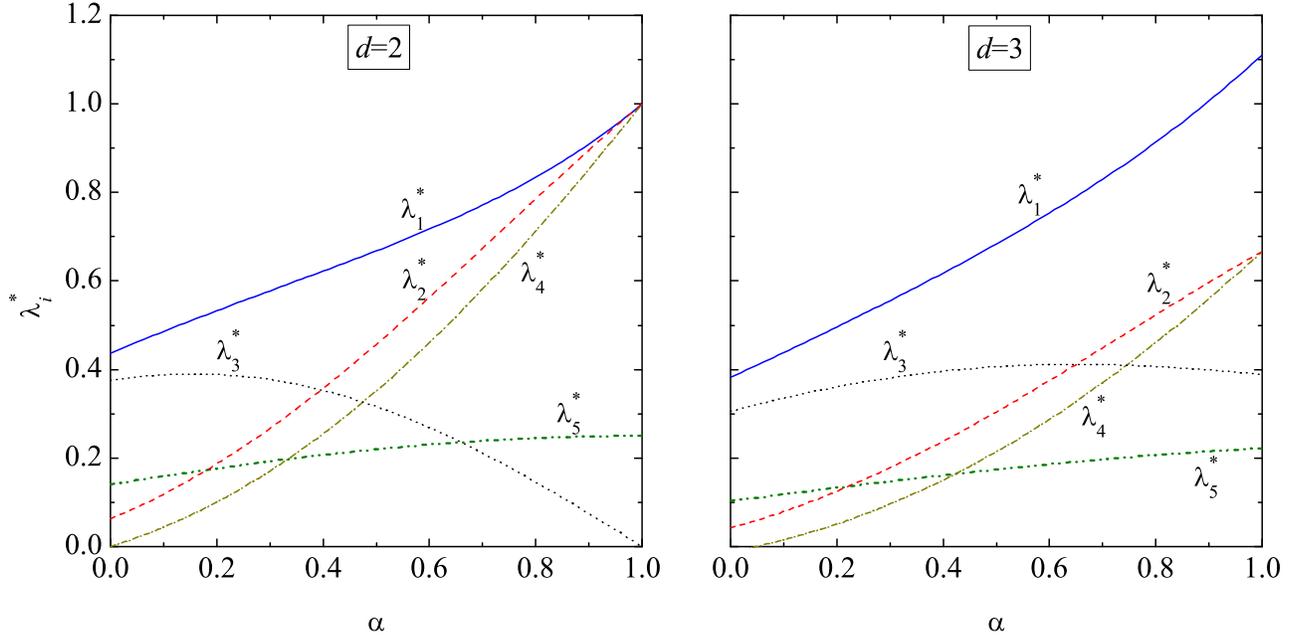}
\caption{(Color online) Plot of the (reduced) cross coefficients
$\lambda_i^*$,
 as functions
of the coefficient of restitution. The left and right panels
correspond to $d=2$ and $d=3$, respectively.
\label{fig2}}
\end{figure}

It is instructive to compare the fourth degree eigenvalues with the
second and third degree eigenvalues. In the elastic case, one has
$\omegafz=\omegato<\omegazt=\omegatt<\omegazth<\omegazf$ for $d=2$
and $\omegafz=\omegato<\omegazt<\omegatt<\omegazth<\omegazf$ for
$d=3$. We have observed that inelasticity breaks the degeneracy
$\omegafz=\omegato$ for both dimensionalities (yielding
$\omegafz<\omegato$) and the degeneracy $\omegazt=\omegatt$ for
$d=2$ (yielding $\omegatt<\omegazt$). The inelasticity also affects
the ordering of the eigenvalues: for $d=2$ one has
$\omegafz<\omegato<\omegatt<\omegazt<\omegazth<\omegazf$  if
$0.17<\alpha<1$ and
$\omegafz<\omegato<\omegatt<\omegazt<\omegazf<\omegazth$  if
$0<\alpha<0.17$; for $d=3$ the ordering is
$\omegafz<\omegato<\omegazt<\omegatt<\omegazth<\omegazf$ if
$0.43<\alpha<1$ and
$\omegafz<\omegato<\omegatt<\omegazt<\omegazth<\omegazf$ if
$0<\alpha<0.43$. Since, except $\omegafz$, these quantities are
related to moments which vanish in isotropic states, the fact that
$\omegafz$ is the smallest one implies that (as expected on physical
grounds) the characteristic time needed to achieve an isotropic
state is shorter than the one needed to reach the asymptotic state.

Let us consider now the (reduced) cross coefficients
$\lambda_i^*\equiv \lambda_i/\nu_0$ ($i=1,\ldots,5$), which measure
the coupling of the second degree moments to the evolution of the
fourth degree moments. Their dependence on dissipation is shown in
Fig.\ \ref{fig2}. It is apparent that the effect of inelasticity on
$\lambda_i^*$ is more pronounced than on the fourth degree
eigenvalues. For elastic collisions,
$\lambda_3^*=0<\lambda_5^*<\lambda_4^*=\lambda_2^*=\lambda_1^*$ for
$d=2$ and
$\lambda_5^*<\lambda_3^*<\lambda_4^*=\lambda_2^*<\lambda_1^*$ for
$d=3$. This ordering changes with inelasticity. Moreover,
$\lambda_1^*$, $\lambda_2^*$, and $\lambda_4^*$ significantly
decrease with increasing dissipation, $\lambda_3^*$ has a
non-monotonic behavior, and $\lambda_5^*$ is nearly constant. Note
that the coefficient $\lambda_4$ does not actually play any role in
$d=2$ since the combination
$M_{0|ik}M_{0|kj}-\frac{1}{d}M_{0|k\ell}M_{0|\ell k}\delta_{ij}$
appearing in the collisional moment $J_{2|i}$, cf.\ Eq.\
\eqref{X11}, vanishes in the two-dimensional case.

\section{Relaxation to the homogeneous cooling state\label{sec3}}
The results derived in the preceding Section can be applied to several interesting situations. Here we will
consider the most basic problem, namely the  time evolution of the moments of degree less than or equal to four
(both isotropic and anisotropic) in the \textit{homogeneous} free cooling state \cite{BP04}. In that case, the
Boltzmann equation \eqref{2.1} becomes
\beq
\partial_t f(\mathbf{v},t)=J[\mathbf{v}|f,f],
\label{Z1} \eeq which must be complemented with a  given initial condition $f(\mathbf{v},0)$. Since the
collisions are inelastic, the granular temperature $T(t)$ monotonically decays in time and so a steady state
does not exist. In the context of IMM, it has been proven \cite{BC03,BCT03} that, provided that
$f(\mathbf{v},0)$ has a finite moment of some degree higher than two, $f(\mathbf{v},t)$ asymptotically tends
toward a self-similar solution of the form \beq f(\mathbf{v},t)\to n [v_0(t)]^{-d} \phi(V/v_0(t)), \label{3.1}
\eeq where $v_0(t)\equiv \sqrt{2T(t)/m}$ is the thermal speed and $\phi(c)$ is an  isotropic distribution that
is only known in the one-dimensional case \cite{BMP02}. According to Eq.\ \eqref{3.1}, the \emph{scaled} moments
\beq
M_{2r|\bar{s}}^*(t)\equiv n^{-1}[v_0(t)]^{-(2r+s)}M_{2r|\bar{s}}(t) \label{3.1a}
\eeq
must tend asymptotically to
\beq M_{2r|\bar{s}}^*(t)\to \mu_{2r|\bar{s}}\equiv \int\dd\mathbf{c}\, Y_{2r|\bar{s}}(\mathbf{c})\phi(c).
\label{3.2}
\eeq
Due to the isotropy of $\phi(c)$, then $\mu_{2r|\bar{s}}=0$ unless $s=0$. Moreover, it is known that the scaled
distribution $\phi(c)$ exhibits an algebraic high velocity tail \cite{EB02a,EB02,KN02,KN02bis} of the form
$\phi(c)\sim c^{-d-\gamma(\alpha)}$, so that the moments $\mu_{2r|0}$ diverge if $2r\geq \gamma(\alpha)$. The
quantity $\gamma(\alpha)$ obeys a transcendental equation whose solution is always $\gamma(\alpha)>4$, except
for $d=1$. Consequently, for any value of $\alpha$ and $d\geq 2$, the scaled moment $M_{4|0}^*(t)$ goes to a
well defined value $\mu_{4|0}$, while the remaining scaled moments of degree equal to or less than four are
anisotropic (except of course $M_{0|0}^*=1$ and $M_{2|0}^*=d/2$) and so they tend to zero. The main goal of this
Section is to analyze in detail the relaxation of the second, third, and fourth degree moments (both isotropic
and anisotropic) toward their asymptotic values.

Taking velocity moments in both sides of Eq.\ \eqref{Z1} one has
\beq
\partial_t M_{2r|\bar{s}}=J_{2r|\bar{s}}.
\label{Z2}
\eeq
In particular,
\beq
\partial_t M_{2|0}=-\nutz M_{2|0}.
\label{Z3} \eeq
Since $M_{2|0}=dnT/m$, Eq.\ \eqref{Z3}  is the equation for the time evolution of the granular
temperature and $\nutz$ is the cooling rate. The solution of Eq.\ \eqref{Z3} is \beq
T(t)=\frac{T(0)}{\left[1+\nutz(0)t/2\right]^2}, \label{Z3b} \eeq where $T(0)$ is the initial temperature and
$\nutz(0)\propto T^{1/2}(0)$ is the initial cooling rate. Equation \eqref{Z3b} is not but Haff's law
\cite{BP04}.

Let us consider now the scaled moments \eqref{3.1a}. In that case,
from Eqs.\ (\ref{Z2}) and (\ref{Z3}) one simply gets
\beq
\partial_\tau M_{2r|\bar{s}}^*=J_{2r|\bar{s}}^*+\frac{2r+s}{2}\nutz^* M_{2r|\bar{s}}^*,
\label{Z5}
\eeq
where  $J_{2r|\bar{s}}^*\equiv J_{2r|\bar{s}}/\nu_0nv_0^{2r+s}$ and
\beq
\tau=\int_0^t\dd t'\, \nu_0(t')
\label{Z6}
\eeq
measures time as the number of (effective) collisions per particle.
The effect of the second term on the right-hand side of Eq.\
\eqref{Z5} is to shift the eigenvalues $\nu_{2r|s}^*$ to
$\omega_{2r|s}=\nu_{2r|s}^*-(r+s/2)\nu_{2|0}^*$. For instance,
\beq
\partial_\tau M_{4|0}^*=-\omegafz M_{4|0}^*+\lambda_1^*
\frac{d^2}{4}-\lambda_2^* M_{0|ij}^*M_{0|ji}^*.
\label{Z8}
\eeq

\subsection{The one-dimensional case}
In the one-dimensional case, the only scaled moments of degree equal
to or less than four are (apart from $M_{0|0}^*=1$ and
$M_{2|0}^*=\frac{1}{2}$) $M_{2|x}^*$ and $M_{4|0}^*$. Their
evolution equations are
\beq
\partial_\tau M_{2|x}^*=0,
\label{Z17}
\eeq
\beq
\partial_\tau
M_{4|0}^*=\frac{3}{16}(1-\alpha^2)^2\left(M_{4|0}^*+\frac{3}{4}\right). \label{Z16} \eeq The solution of Eq.\
\eqref{Z16} is \beq M_{4|0}^*(\tau)=M_{4|0}^*(0)
e^{\frac{3}{16}(1-\alpha^2)^2\tau}+\frac{3}{4}\left[e^{\frac{3}{16}(1-\alpha^2)^2\tau}-1\right]. \label{Z18.2}
\eeq
This solution shows that the (scaled) fourth degree moment monotonically increases with time, i.e.,
$\mu_{4|0}=\infty$. This is consistent with the exact HCS solution found by Baldassarri \emph{et
al}.\cite{BMP02}, namely
\beq \phi(c)=\frac{2^{3/2}}{\pi}\frac{1}{(1+2c^2)^2}. \label{Z18}
\eeq

On the other hand, Eq.\ \eqref{Z17} shows that
$M_{2|x}^*(\tau)=M_{2|x}^*(0)$, i.e., if the initial state is
anisotropic with $M_{2|x}(0)\neq 0$ then one has
$M_{2|x}(t)=M_{2|x}(0)[T(t)/T(0)]^{3/2}$. The constancy of
$M_{2|x}^*$ implies that any initial anisotropy does not vanish in
the scaled velocity distribution function for long times. As a
consequence, while the distribution \eqref{Z18} represents the
asymptotic form $\phi(c)$ for a wide class of \emph{isotropic}
initial conditions, it cannot be reached, strictly speaking, for any
\emph{anisotropic} initial state. Whether or not there exists a
generalization of \eqref{Z18} for anisotropic states is, to the best
of our knowledge, an open problem. Since the symmetry of the
distribution \eqref{Z18} implies that $M_{2|x}^*\equiv\langle
c_x^3\rangle =0$ but the average $\langle |c|^3\rangle$ diverges,  a
small correction to the form \eqref{Z18} could accommodate a finite
value of $ \langle c_x^3\rangle$.

\subsection{The two-dimensional case}
As is known, one-dimensional systems are generally not very
realistic and so they can exhibit peculiar properties. However,
two-dimensional systems are usually considered as representative of
the features present in real systems.

For the sake of simplicity, in the remainder of this Section we will consider the two-dimensional case. The set
of independent moments of second, third, and fourth degree will be taken as
\beq \{M_{0|xx}^*,M_{0|xy}^*\},
\eeq \beq \{M_{2|x}^*,M_{2|y}^*,M_{0|xxx}^*,M_{0|xxy}^*\},
\eeq
\beq
\{M_{4|0}^*,M_{2|xx}^*,M_{2|xy}^*,M_{0|xxxx}^*,M_{0|xxxy}^*\}, \label{4th}
\eeq
respectively. The remaining moments are simply related to the above ones as $M_{0|yy}^*=-M_{0|xx}^*$,
$M_{0|xyy}^*=-M_{0|xxx}^*$, $M_{0|yyy}^*=-M_{0|xxy}^*$,  $M_{2|yy}^*=-M_{2|xx}^*$,
$M_{0|yyyy}^*=-M_{0|xxyy}^*=M_{0|xxxx}^*$, $M_{0|xyyy}^*=-M_{0|xxxy}^*$. {}From Eq.\ \eqref{Z5}, it is easy to
obtain the time dependence of the (scaled) second and third degree moments:
\beq
M_{0|ij}^*(\tau)=M_{0|ij}^*(0)e^{-\omegazt \tau}, \label{Z13}
\eeq
\beq M_{2|i}^*(\tau)=M_{2|i}^*(0)e^{-\omegato
\tau}, \quad M_{0|ijk}^*(\tau)=M_{0|ijk}^*(0)e^{-\omegazth \tau}.
\eeq
In the case of the fourth degree moments, one has to deal with inhomogeneous linear differential equations
involving the second degree moments. The solutions are
\beqa M_{4|0}^*(\tau)&=&M_{4|0}^*(0)e^{-\omegafz
\tau}+\frac{\lambda_1^*}{\omegafz}\left(1-e^{-\omegafz \tau}\right)\nn
&&-\frac{2\lambda_2^*}{2\omegazt-\omegafz}\left[M_{0|xx}^{*2}(0)+M_{0|xy}^{*2}(0)\right]\left(e^{-\omegafz
\tau}-e^{-2\omegazt \tau}\right), \label{Z9}
\eeqa
\beq M_{2|ij}^*(\tau)=M_{2|ij}^*(0)e^{-\omegatt
\tau}+\frac{\lambda_3^*}{\omegazt-\omegatt}M_{0|ij}^*(0)\left(e^{-\omegatt \tau}-e^{-\omegazt \tau}\right),
\label{Z10} \eeq \beq M_{0|xxxx}^*(\tau)=M_{0|xxxx}^*(0)e^{-\omegazf
\tau}+\frac{3}{2}\frac{\lambda_5^*}{2\omegazt-\omegazf}\left[M_{0|xx}^{*2}(0)-M_{0|xy}^{*2}(0)\right]\left(e^{-\omegazf
\tau}-e^{-2\omegazt \tau}\right), \label{Z11}
\eeq
\beq M_{0|xxxy}^*(\tau)=M_{0|xxxy}^*(0)e^{-\omegazf
\tau}+\frac{3\lambda_5^*}{2\omegazt-\omegazf}M_{0|xx}^{*}(0)M_{0|xy}^{*}(0)\left(e^{-\omegazf
\tau}-e^{-2\omegazt \tau}\right). \label{Z12}
\eeq
Equations \eqref{Z13}--\eqref{Z12} show that all the moments,
except $M_{4|0}^*$, tend to zero for sufficiently long times. The asymptotic expression of $M_{4|0}^*$ is
\beq
M_{4|0}^*\to\mu_{4|0}=\frac{\lambda_1^*}{\omegafz}=2\frac{7-6\alpha+3\alpha^2}{1+6\alpha-3\alpha^2}, \label{Z14}
\eeq
which agrees with previous results \cite{S03}.

\begin{figure}
\includegraphics[width=0.95\columnwidth]{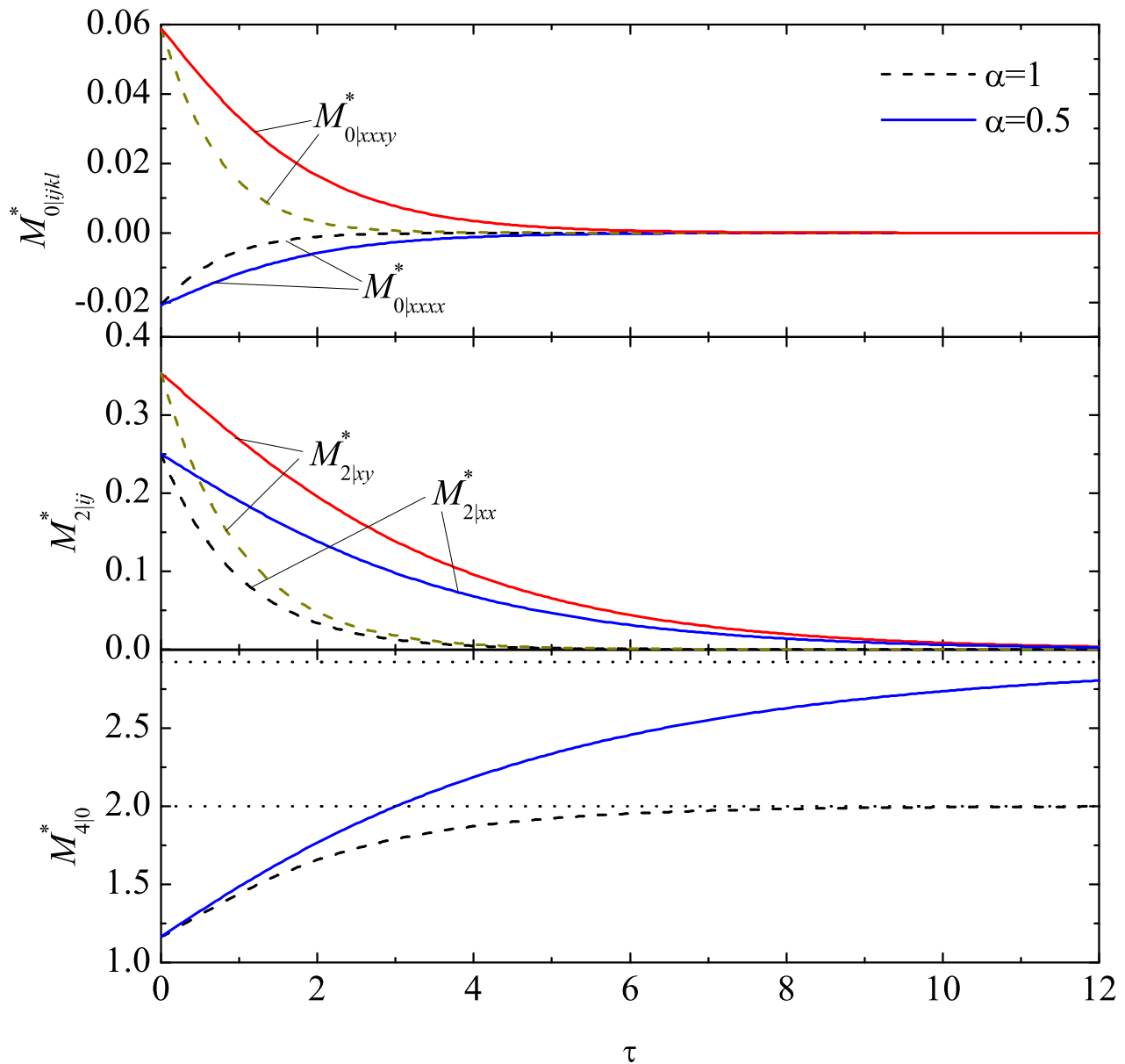}
\caption{(Color online) Time evolution of the scaled fourth degree
moments in the free cooling of a two-dimensional IMM with
$\alpha=0.5$ (solid lines). The time evolution to equilibrium of an
elastic system ($\alpha=1$) is also shown (dashed lines). In both
cases the initial state is described by Eq.\ \protect\eqref{Z15}.
The horizontal dotted lines in the bottom panel indicate the
corresponding asymptotic HCS values $\mu_{4|0}$.
\label{fig3}}
\end{figure}

As an illustration, let us analyze the time evolution of the scaled
fourth degree moments \eqref{4th} for the following initial
anisotropic distribution:
\beq
f(\mathbf{v},0)=\frac{n}{3}\left[\delta(\mathbf{V}-\mathbf{V}_1)+\delta(\mathbf{V}-\mathbf{V}_2)+\delta(\mathbf{V}-\mathbf{V}_3)\right],
\label{Z15}
\eeq
where $\mathbf{V}_1=v_0(0)\widehat{\mathbf{x}}$,
$\mathbf{V}_2=(v_0(0)/\sqrt{2})\widehat{\mathbf{y}}$, and
$\mathbf{V}_3=-\mathbf{V}_1-\mathbf{V}_2$. Here,
$v_0(0)=\sqrt{2T(0)/m}$ is the initial thermal speed, where the
initial temperature $T(0)$ is arbitrary. Figure \ref{fig3} shows the
evolution of the moments \eqref{4th} for two values of the
coefficient of restitution: $\alpha=1$ (elastic system) and
$\alpha=0.5$ (strongly inelastic system). It is quite apparent that
the number of collisions needed to reach the HCS values increases
with the inelasticity, as expected from Fig.\ \ref{fig1}. In the
particular case of $\alpha=0.5$, the relaxation times are about
twice the ones corresponding to $\alpha=1$. Moreover, since
$\omegazf>\omegatt>\omegafz$, we observe that the moments
$M_{0|ijk\ell}^*$ tend to zero more rapidly than the moments
$M_{2|ij}^*$, and that the isotropic moment $M_{4|0}^*$ reaches its
asymptotic value more slowly than the anisotropic moments.

\section{Discussion\label{sec4}}
As Maxwell already realized \cite{Maxwell}, scattering models where the collision rate of two particles
approaching each other with a relative velocity $\mathbf{g}$ is independent of the magnitude of $\mathbf{g}$
allows one to evaluate exactly the collisional moments without the explicit knowledge of the velocity
distribution function. In the conventional case of ordinary gases of particles colliding elastically, Maxwell
models are useful to find non-trivial exact solutions to the Boltzmann equation in far from equilibrium
situations \cite{GS03}. Needless to say, the introduction of inelasticity through a constant coefficient of
normal restitution $\alpha\leq 1$ opens up new perspectives for exact results, including the elastic case as a
special limit ($\alpha=1$). This justifies the growing interest in IMM by physicists and mathematicians alike in
the past few years.

The choice of the Ikenberry polynomials \cite{TM80} $Y_{2r|\bar{s}}$ of degree $2r+s$ allows one to express the
corresponding collisional moments $J_{2r|\bar{s}}$ in the form \eqref{2.4}: an eigenvalue $-\nu_{2r|s}$ times
the velocity moment $M_{2r|\bar{s}}$ plus a bilinear combination of moments of degree less than $2r+s$. In
particular, $\nu_{2|0}$ is the cooling rate of the gas. In this paper we have evaluated all the third and fourth
degree collisional moments of the IMM defined by Eq.\ \eqref{1}. In that context, the results are exact for
arbitrary values of $\alpha$ and apply to any dimensionality $d$. Known results are recovered for
three-dimensional elastic systems \cite{TM80,GS03} and for one-dimensional inelastic systems \cite{NK00}. We
have observed that some of the eigenvalues $\nu_{2r|s}$ do not have a monotonic dependence on $\alpha$, while
the shifted eigenvalues $\nu_{2r|s}-(r+s/2)\nu_{2|0}$ monotonically decrease with increasing inelasticity.
Moreover, given a value of $\alpha$ and a degree $2r+s$, the eigenvalues $\nu_{2r|s}$ increase with $s$. This
means that the larger the anisotropy of a moment $M_{2r|\bar{s}}$ the higher its collisional rate of change.
Although the above observations are based on the moments of degree $2r+s\leq 4$, we expect that they extend to
moments of higher degree.

As a simple application of the results derived in Section
\ref{sec2}, we have studied the time evolution of the  moments of
degree equal to or less than four in the free cooling state, in
which case the evolution of the moments scaled with the thermal
speed is governed by the shifted eigenvalues. An interesting feature
of the one-dimensional case is that  the heat flux
$q_x=(m/2)M_{2|x}$, when scaled with the thermal speed, does not
change in time, so that an initial anisotropic distribution cannot
evolve toward an asymptotic isotropic distribution. Thus, the exact
solution found by Baldassarri \emph{et al.}\cite{BMP02} does not
play a universal role, at least in a strict sense, unless the
initial distribution is isotropic. On the other hand, we have found
that all the anisotropic moments of degree equal to or less than
four vanish in the long time limit for $d\geq 2$. However, this does
not preclude the possibility that anisotropic moments of higher
degree diverge for $\alpha$ sufficiently small. We plan to explore
this possibility in the near future.

The explicit results provided in this paper can be useful for
studying different problems. An important application is the exact
derivation of the Burnett order constitutive equations for IMM, with
explicit expressions of the associated transport coefficients as
functions of $d$ and $\alpha$. This is possible because the
determination of the Burnett order pressure tensor and heat flux
requires the previous knowledge of the third and fourth degree
 collisional moments to Navier--Stokes order. Another
interesting problem is the so-called simple or uniform shear flow, which is an intrinsically non-Newtonian state
\cite{SGD04}. Apart from the rheological quantities, the results derived here allows one to analyze the time
evolution of the fourth degree velocity moments toward their steady state values \cite{SG07} and investigate
their possible divergence, in a similar way to the analysis carried out in the elastic case \cite{SGBD93}.
Finally, the generalized transport coefficients characterizing small perturbations around the simple shear flow
have been determined \cite{G07} and compared with those previously obtained for IHS \cite{G06} from a model
kinetic equation.

\acknowledgments

This research  has been supported by the Ministerio de Educaci\'on y Ciencia (Spain) through Grant No.\
FIS2007-60977, partially financed by FEDER funds.

\appendix*
\section{Explicit evaluation of the collisional moments\label{appA}}
In this Appendix we give the details of the derivation,   by using
the property \eqref{5}, of the collisional moments
$J_{2r|\bar{s}}=\mathcal{J}[Y_{2r|\bar{s}}]$ associated with the
Ikenberry polynomials of third and fourth degree. To carry out the
calculations we will need the angular integrals
\begin{equation}
\label{N9}
\int
\dd\widehat{\boldsymbol{\sigma}}\,(\widehat{\boldsymbol{\sigma}}\cdot
{\bf g})^{2\qq+1} \widehat{{\sigma}}_i=B_{\qq+1} g^{2\qq}{g_i},
\end{equation}
\begin{equation}
\label{N10}
\int
\dd\widehat{\boldsymbol{\sigma}}\,(\widehat{\boldsymbol{\sigma}}\cdot
{\bf g})^{2\qq} \widehat{{\sigma}}_i\widehat{{\sigma}}_j
=\frac{B_{\qq}}{2\qq+d} g^{2(\qq-1)}\left(2\qq{g_i}{g_j}+g^2
\delta_{ij}\right),
\end{equation}
\begin{equation}
\label{N21}
\int
\dd\widehat{\boldsymbol{\sigma}}\,(\widehat{\boldsymbol{\sigma}}\cdot
{\bf g})^{2\qq+1}
\widehat{\sigma}_i\widehat{\sigma}_j\widehat{\sigma}_k=\frac{B_{\qq+1}}{2(\qq+1)+d}
g^{2(\qq-1)}\left[ 2\qq
g_ig_jg_k+g^2\left(\delta_{ij}g_k+\delta_{ik}g_j+\delta_{jk}g_i\right)\right],
\end{equation}
\begin{eqnarray}
\label{N29}
\int
\dd\widehat{\boldsymbol{\sigma}}\,(\widehat{\boldsymbol{\sigma}}\cdot
{\bf g})^{2\qq}
\widehat{\sigma}_i\widehat{\sigma}_j\widehat{\sigma}_k
\widehat{\sigma}_{\ell}&=&\frac{B_{\qq}}{(2\qq+d)[d+2(\qq+1)]}
\left[4\qq(\qq-1)g^{2(\qq-2)}g_ig_jg_kg_{\ell}\right. \nonumber\\
& &
+g^{2(\qq-1)}2\qq\left(g_ig_j\delta_{k\ell}+g_ig_k\delta_{j\ell}+g_ig_{\ell}\delta_{jk}+
g_kg_j\delta_{i\ell}+g_{\ell}g_j\delta_{ik}+g_kg_{\ell}\delta_{ij}\right)\nonumber\\
& & \left.
+g^{2\qq}\left(\delta_{ij}\delta_{k\ell}+\delta_{ik}\delta_{j\ell}+\delta_{jk}\delta_{i\ell}\right)
\right].
\end{eqnarray}
Here, the coefficients $B_\qq$ are \cite{NE98}
\begin{equation}
\label{N11}
B_{\qq}=\int
\dd\widehat{\boldsymbol{\sigma}}\,(\widehat{\boldsymbol{\sigma}}\cdot
{\widehat{\bf
g}})^{2\qq}=\Omega_d\pi^{-1/2}\frac{\Gamma\left(\frac{d}{2}\right)
\Gamma\left(\qq+\frac{1}{2}\right)}{\Gamma\left(\qq+\frac{d}{2}\right)}.
\end{equation}

\subsection{Third degree moments\label{appAbis}}
We start by noting that the collision rule \eqref{6} implies that
\beqa
V_{1i}''V_{1j}''V_{1k}''&=&V_{1i}V_{1j}V_{1k}-
\at\sg\left(V_{1i}V_{1j}\si_k+V_{1i}V_{1k}\si_j+V_{1j}V_{1k}\si_i\right)\nn
&&+
\left(\at\right)^2\sg^2\left(V_{1i}\si_j\si_k+V_{1j}\si_i\si_k+V_{1k}\si_i\si_j\right)-
\left(\at\right)^3\sg^3\si_i\si_j\si_k.\nn &&
\label{A1}
\eeqa
Next, making use of Eqs.\ \eqref{N9}--\eqref{N21}, one gets
\beqa
\mathcal{J}[V_{1i}V_{1j}V_{1k}]&=&-\frac{n\nu_0}{2d}\at\left\{(d+2)\langle
g_iV_{1j}V_{1k}+g_jV_{1i}V_{1k}+g_kV_{1i}V_{1j}\rangle\right.\nn
&&\left.-\at\left[2\langle
g_ig_jV_{1k}+g_ig_kV_{1j}+g_jg_kV_{1i}\rangle+\langle
g^2\left(V_{1i}\delta_{jk}+V_{1j}\delta_{ik}+V_{1k}\delta_{ij}\right)\rangle\right]\right\},\nn
&&
\label{A2}
\eeqa
where the brackets are defined as
\begin{equation}
\label{N13}
\langle h({\bf V}_1,{\bf V}_2)\rangle \equiv \frac{1}{n^2}\int
\dd{\bf V}_1\int \dd{\bf V}_2 h({\bf V}_1,{\bf V}_2) f({\bf
V}_1)f({\bf V}_2).
\end{equation}
 and we have taken
into account that $\langle g_ig_jg_k\rangle=\langle
g^2g_{i}\rangle=0$. It is easy to get
\beq
n\langle g_iV_{1j}V_{1k}\rangle=n\langle
g_ig_{j}V_{1k}\rangle=M_{0|ijk}+\frac{1}{d+2}\left(M_{2|i}\delta_{jk}+M_{2|j}\delta_{ik}+M_{2|k}\delta_{ij}\right),
\label{A3}
\eeq
\beq
n\langle g^2V_{1i}\rangle=M_{2|i}.
\label{A4}
\eeq
Therefore,
\beq
\mathcal{J}[V_{1i}V_{1j}V_{1k}]=-\frac{\nu_0}{2d}\at\left[3(d+1-\alpha)M_{0|ijk}+\frac{5d+4-\alpha(d+8)}{2(d+2)}
\left(M_{2|i}\delta_{jk}+M_{2|j}\delta_{ik}+M_{2|k}\delta_{ij}\right)\right].
\label{A5}
\eeq
If one makes $j=k$ and sum over $j$ one gets the first equality of
Eq.\ (\ref{X7}) with $\nu_{2|1}$ given by  Eq.\ (\ref{X7b}). Also,
by subtracting
$\left(J_{2|i}\delta_{jk}+J_{2|j}\delta_{ik}+J_{2|k}\delta_{ij}\right)/(d+2)$
from both sides of Eq.\ (\ref{A5}) one gets the second equality of
Eq.\ (\ref{X7}) with $\nu_{0|3}$ given by  Eq.\ (\ref{X8}).

\subsection{Fourth degree moments\label{appB}}
Now the starting point is the collision rule
\beqa
V_{1i}''V_{1j}''V_{1k}''V_{1\ell}''&=&V_{1i}V_{1j}V_{1k}V_{1\ell}-
\at\sg\left(V_{1i}V_{1j}V_{1k}\si_\ell+V_{1i}V_{1j}V_{1\ell}\si_k+V_{1i}V_{1k}V_{1\ell}\si_j+V_{1j}V_{1k}V_{1\ell}\si_i\right)\nn
&&+
\left(\at\right)^2\sg^2\left(V_{1i}V_{1j}\si_k\si_\ell+V_{1i}V_{1k}\si_j\si_\ell+V_{1i}V_{1\ell}\si_j\si_k
+V_{1j}V_{1k}\si_i\si_\ell+V_{1j}V_{1\ell}\si_i\si_k\right.\nn
&&\left.+V_{1k}V_{1\ell}\si_i\si_j\right)-
\left(\at\right)^3\sg^3\left(V_{1i}\si_j\si_k\si_\ell+V_{1j}\si_i\si_k\si_\ell+V_{1k}\si_i\si_j\si_\ell+V_{1\ell}\si_i\si_j\si_k\right)\nn
&&+\atp^4\sg^4\si_i\si_j\si_k\si_\ell.
\label{B1}
\eeqa
After integrating over $\widehat{\bm{\sigma}}$,
\beqa
\mathcal{J}[V_{1i}V_{1j}V_{1k}V_{i\ell}]&=&-\frac{n\nu_0}{2d(d+4)(d+6)}\at\left\{(d+2)(d+4)(d+6)\langle
g_iV_{1j}V_{1k}V_{1\ell}+\stackrel{(3)}{\cdots}\rangle\right.\nn
&&-\at(d+4)(d+6)\left[2\langle
g_ig_jV_{1k}V_{1\ell}+\stackrel{(5)}{\cdots}\rangle+\langle
g^2(V_{1i}V_{ij}\delta_{k\ell}+\stackrel{(5)}{\cdots})\rangle\right]\nn
&& +3(d+6)\atp^2 \left[2\langle
g_ig_jg_kV_{1\ell}+\stackrel{(3)}{\cdots}\rangle+\langle
g^2[(g_iV_{1j}+g_jV_{1i})\delta_{k\ell}+\stackrel{(5)}{\cdots}]\rangle\right]\nn
&& \left. -3\atp^3\left[8\langle g_ig_jg_kg_\ell\rangle+4\langle
g^2(g_ig_j\delta_{k\ell}+\stackrel{(5)}{\cdots})\rangle+\langle
g^4\rangle
(\delta_{ij}\delta_{k\ell}+\stackrel{(2)}{\cdots})\right]\right\},\nn
&&
\label{B2}
\eeqa
where $\stackrel{(s)}{\cdots}$ denotes the $s$ terms obtained from
the canonical one by permutation of indices. Making $k=\ell$ and
summing over $k$ we obtain
\beqa
\mathcal{J}[V_1^2V_{1i}V_{1j}]&=&-\frac{n\nu_0}{2d(d+4)}\at\left\{(d+2)(d+4)\langle
2 (\mathbf{g}\cdot\mathbf{V}_1)V_{1i}V_{1j}+V^2(
g_iV_{1j}+g_jV_{1i})\rangle\right.\nn &&-\at(d+4)\left[2\langle
V_1^2g_ig_j+2(\mathbf{g}\cdot\mathbf{V}_1)(g_iV_{1j}+g_j
V_{1i})\rangle +\langle
(d+6)g^2V_{1i}V_{1j}+g^2V_1^2\delta_{ij}\rangle\right]\nn &&
+3\atp^2 \left[(d+6)\langle g^2(g_iV_{1j}+g_j
V_{1i})\rangle+2\langle
(\mathbf{g}\cdot\mathbf{V}_1)(2g_ig_j+g^2\delta_{ij})\rangle\right]\nn
&& \left. -3\atp^3\langle 4g^2g_ig_j+g^4\delta_{ij}\rangle\right\}.
\label{B2.2}
\eeqa
 Next, by
taking $i=j$ and summing over $i$, Eq.\ (\ref{B2.2}) yields
\beqa
\mathcal{J}[V_1^4]&=&-\frac{n\nu_0}{2d}\at\left\{4(d+2)\langle
(\mathbf{g}\cdot\mathbf{V}_1)V_1^2\rangle-2\at\left[(d+4)\langle
g^2V_1^2\rangle +4\langle
(\mathbf{g}\cdot\mathbf{V}_1)^2\rangle\right]\right.\nn  &&\left.
+12\atp^2 \langle (\mathbf{g}\cdot\mathbf{V}_1)g^2\rangle
-3\atp^3\langle g^4\rangle\right\}.
\label{B2.3}
\eeqa

Now we express the averages in terms of the Ikenberry moments. Let
us consider first the four-index averages:
\beq
n\langle
g_iV_{1j}V_{1k}V_{1\ell}\rangle=M_{0|ijk\ell}+\frac{1}{d+4}\left(M_{2|ij}\delta_{k\ell}+\stackrel{(5)}{\cdots}\right)+\frac{1}{d(d+2)}
M_{4|0}\left(\delta_{ij}\delta_{k\ell}+\stackrel{(2)}{\cdots}\right),
\label{B3}
\eeq
\beqa
n\langle
g_ig_{j}V_{1k}V_{1\ell}\rangle&=&M_{0|ijk\ell}+\frac{1}{d+4}\left(M_{2|ij}\delta_{k\ell}+\stackrel{(5)}{\cdots}\right)+\frac{1}{d(d+2)}
M_{4|0}\left(\delta_{ij}\delta_{k\ell}+\stackrel{(2)}{\cdots}\right)\nn
&&
+n^{-1}M_{0|ij}M_{0|k\ell}+\frac{n^{-1}}{d}M_{2|0}\left(M_{0|ij}\delta_{k\ell}+M_{0|k\ell}\delta_{ij}\right)
+\frac{n^{-1}}{d^2}M_{2|0}^2\delta_{ij}\delta_{k\ell},\nn &&
\label{B4}
\eeqa
\beqa
n\langle
g_ig_{j}g_{k}V_{1\ell}\rangle&=&M_{0|ijk\ell}+\frac{1}{d+4}\left(M_{2|ij}\delta_{k\ell}+\stackrel{(5)}{\cdots}\right)+\frac{1}{d(d+2)}
M_{4|0}\left(\delta_{ij}\delta_{k\ell}+\stackrel{(2)}{\cdots}\right)\nn
&&
+n^{-1}\left(M_{0|ij}M_{0|k\ell}+\stackrel{(2)}{\cdots}\right)+\frac{n^{-1}}{d}M_{2|0}\left(M_{0|ij}\delta_{k\ell}+\stackrel{(5)}{\cdots}\right)
+\frac{n^{-1}}{d^2}M_{2|0}^2\left(\delta_{ij}\delta_{k\ell}+\stackrel{(2)}{\cdots}\right),\nn
&&
\label{B5}
\eeqa
\beq
\langle g_ig_{j}g_{k}g_{\ell}\rangle=2\langle
g_ig_{j}g_{k}V_{1\ell}\rangle.
\label{B6}
\eeq
Summing over two repeated indices we get the two-index averages:
\beq
n\langle g^2V_{1i}V_{1j}\rangle=n\langle
V_1^2g_ig_j\rangle=M_{2|ij}+\frac{1}{d}M_{4|0}\delta_{ij}+n^{-1}M_{2|0}\left(M_{0|ij}+\frac{1}{d}M_{2|0}\delta_{ij}\right),
\label{B7}
\eeq
\beqa
n\langle g^2g_{i}V_{1j}\rangle&=&n\langle
(\mathbf{g}\cdot\mathbf{V}_1)g_{i}g_{j}\rangle=M_{2|ij}+\frac{1}{d}M_{4|0}\delta_{ij}+2n^{-1}M_{0|ik}M_{0|kj}\nn
&&+\frac{n^{-1}}{d}M_{2|0}\left[(d+4)M_{0|ij}
+\frac{d+2}{d}M_{2|0}\delta_{ij}\right],
\label{B8}
\eeqa
\beq
\langle g^2g_{i}g_{j}\rangle=2\langle g^2g_{i}V_{1j}\rangle,
\label{B9}
\eeq
\beq
n\langle V_1^2 g_iV_{1j}\rangle=n\langle
(\mathbf{g}\cdot\mathbf{V}_1)V_{1i}V_{1j}\rangle=M_{2|ij}+\frac{1}{d}M_{4|0}\delta_{ij},
\label{B11}
\eeq
\beq
n\langle
(\mathbf{g}\cdot\mathbf{V}_1)g_{i}V_{1j}\rangle=M_{2|ij}+\frac{1}{d}M_{4|0}\delta_{ij}+n^{-1}M_{0|ik}M_{0|kj}+\frac{n^{-1}}{d}M_{2|0}\left[
2M_{0|ij}+\frac{1}{d}M_{2|0}\delta_{ij}\right].
\label{B13}
\eeq
Summing again,
\beq
n\langle g^2V_1^2\rangle=M_{4|0}+n^{-1}M_{2|0}^2,
\label{B12}
\eeq
\beq
n\langle (\mathbf{g}\cdot\mathbf{V}_1)V_1^2\rangle=M_{4|0},
\label{B17}
\eeq
\beq
n\langle
(\mathbf{g}\cdot\mathbf{V}_1)^2\rangle=M_{4|0}+n^{-1}M_{0|ij}M_{0|ji}+\frac{n^{-1}}{d}M_{2|0}^2,
\label{B15}
\eeq
\beq
n\langle
(\mathbf{g}\cdot\mathbf{V}_1)g^2\rangle=M_{4|0}+2n^{-1}M_{0|ij}M_{0|ji}+n^{-1}\frac{d+2}{d}M_{2|0}^2,
\label{B16}
\eeq
\beq
\langle g^4\rangle=2\langle (\mathbf{g}\cdot\mathbf{V}_1)g^2\rangle.
\label{B10}
\eeq

Inserting Eqs.\ (\ref{B12})--(\ref{B10}) into Eq.\ (\ref{B2.3}) one
gets Eq.\ (\ref{X9}). Analogously, from Eqs.\ (\ref{B2.2}) and
(\ref{B2}) one gets, after some algebra, Eqs.\ (\ref{X11}) and
(\ref{X12}), respectively.

\end{document}